\documentclass[aps, prd, showpacs,floatfix, superscriptaddress]{revtex4-2}
\usepackage{amsmath,amssymb,amsfonts}
\usepackage{float}
\usepackage{graphicx}
\usepackage{xcolor}
\usepackage{tensor}
\usepackage{lineno}
\usepackage[colorlinks,linkcolor=blue,citecolor=blue,urlcolor=blue]{hyperref}
\usepackage{appendix}
\usepackage{comment}

\newcommand{\be}{\begin{eqnarray}}
	\newcommand{\ee}{\end{eqnarray}}

\begin{document}


\title{Causal third-order viscous hydrodynamics within relaxation-time approximation}

\date{ \today }

\author{Pushpa Panday\footnote{pandaypushpa147@gmail.com}}
\affiliation{Department of Physics, Indian Institute of Technology Roorkee, Roorkee - 247667, India}
 
\author{Amaresh Jaiswal\footnote{a.jaiswal@niser.ac.in}}
\affiliation{School of Physical Sciences, National Institute of Science Education and Research, An OCC of Homi Bhabha National Institute, Jatni-752050, India}
\affiliation{Institute of Theoretical Physics, Jagiellonian University, ul. St. \L ojasiewicza 11, 30-348 Krakow, Poland}

\author{Binoy Krishna Patra\footnote{binoy@ph.iitr.ac.in}}
\affiliation{Department of Physics, Indian Institute of Technology Roorkee, Roorkee - 247667, India}

%
\begin{abstract} 
In the present work, we derive a linearly stable and causal theory of relativistic third-order viscous hydrodynamics from the Boltzmann equation with relaxation-time approximation. We employ viscous correction to the distribution function obtained using a Chapman-Enskog like iterative solution of the Boltzmann equation. Our derivation highlights the necessity of incorporating a new dynamical degree of freedom, specifically an irreducible tensor of rank three, within this framework. This differs from the recent formulation of causal third-order theory from the method of moments which requires two dynamical degrees of freedom: an irreducible third-rank and a fourth-rank tensor. We verify the linear stability and causality of the proposed formulation by examining perturbations around a global equilibrium state.
\end{abstract}
%

\maketitle


\section{Introduction}
\label{intro}

The primary objective of ultra-relativistic heavy-ion collisions is to create and study a novel state of nuclear matter characterized by extremely high temperature and/or density. In this extreme environment, composite states known as hadrons lose their distinct identity, undergoing dissolution into a quark-gluon plasma (QGP), where quarks and gluons exist in a deconfined state~\cite{Collins:1974ky, Baumgardt:1975qv, Shuryak:1978ij, Kapusta:1979fh, Elze:1989un, Rischke:2003mt, Csernai:2006zz}. Relativistic dissipative hydrodynamics has been successfully applied to study the collective behaviour of QGP~\cite{Gale:2013da, Heinz:2013th}. These collisions give rise to a fluid under extreme conditions, characterized by gradients of fluid velocity and temperature that are substantial when compared to the characteristic microscopic scales of the system~\cite{Heller:2015dha, Romatschke:2017vte}. The effort to understand the hydrodynamic behavior of QGP created in ultra-relativistic heavy-ion collisions has spurred significant research in formulation of relativistic dissipative fluid dynamics from microscopic theory—a topic that continues to be an area of active research to this day~\cite{Florkowski:2017olj, Romatschke:2017ejr}.

Relativistic dissipative hydrodynamics is formulated through an order-by-order expansion in powers of spacetime gradients, with ideal hydrodynamics corresponding to the zeroth order. The first order theory containing the viscous effects, commonly referred to as the relativistic Navier-Stokes theory~\cite{Landau, Eckart:1940zz}, is known to be ill-defined since it involves parabolic differential equations which leads to acausality and numerical instability. Causality was restored in second order Israel-Stweart (IS) theory~\cite{Israel:1979wp} with its hyperbolic equations~\cite{Huovinen:2008te} although stability may not be assured. On the other hand, IS theory lead to some undesirable effects such as reheating of the expanding medium~\cite{Muronga:2003ta} and emergence of negative longitudinal pressure~\cite{Martinez:2009mf, Rajagopal:2009yw}. Moreover, the scaling solutions of IS theory exhibit disagreement with transport results for large viscosities, indicating the breakdown of second-order theory. It was argued that an empirical inclusion of higher-order terms significantly improved the agreement with transport result~\cite{El:2009vj, Jaiswal:2013npa}, highlighting the necessity of formulating relativistic dissipative hydrodynamics beyond the second-order IS theory.

Several authors have explored the formulation of relativistic third-order dissipative fluid-dynamics within various frameworks such as phenomenological description based on the second law of thermodynamics~\cite{El:2009vj, Younus:2019rrt} and kinetic theory using Chapman-Enskog-like expansion~\cite{Jaiswal:2013vta} and gradient expansion~\cite{Grozdanov:2015kqa}. Recently, the linear stability and causality of third-order theory, formulated in Ref.~\cite{Jaiswal:2013vta}, was analyzed and was shown to be acausal and unstable~\cite{Brito:2021iqr}. In order to address this issue, a heuristic modification to this theory was proposed by introducing a new dynamical degree of freedom~\cite{Brito:2021iqr}. In a follow-up work, the authors derived this framework from kinetic theory using the method of moments~\cite{deBrito:2023tgb}. This formulation required the inclusion of novel degrees of freedom corresponding to irreducible tensors of rank 3 and 4, and was shown to be in good agreement with exact solution of the Boltzmann equation within Bjorken flow scenario. On the other hand, iterative Chapman-Enskog method has also been employed quite successfully in the formulation of relativistic dissipative hydrodynamics, leading to good agreement with kinetic theory solutions~\cite{Jaiswal:2013vta, Jaiswal:2013npa}. Moreover, it was shown that the non-equilibrium corrections to the distribution function, obtained using maximum entropy prescription, matches exactly with the iterative Chapman-Enskog results at linear order~\cite {Everett:2021ulz, Chattopadhyay:2023hpd}. It is therefore important to explore the formulation of causal and stable third-order hydrodynamics with iterative Chapman-Enskog-like expansion of the Boltzmann equation in relaxation-time approximation.

In this article, we present the derivation of a linearly stable and causal theory of relativistic third-order viscous hydrodynamics from the Boltzmann equation with relaxation-time approximation. To this end, we use viscous corrections to the distribution function obtained from a Chapman-Enskog-like iterative solution of the Boltzmann equation. The derivation underscores the essential inclusion of a novel dynamical degree of freedom in this framework, namely an irreducible three-rank tensor. This is in contrast with the recent formulation of causal third-order theory from the method of moments which required two dynamical degrees of freedom for moment closure: an irreducible third-rank and a fourth-rank tensors~\cite{deBrito:2023tgb}. To validate our formulation, we analyze its linear stability and causality by investigating perturbations around a global equilibrium state. This work is organised as follows: In Sec.~\eqref{3rd-ord}, we review the derivation of third-order viscous evolution equation using Chapman-Enskog like iterative solution of the Boltzmann equation. In Sec.~\ref{3rd-csl}, we derive the framework to restore causality in third-order viscous evolution equation. In Sec.~\eqref{stb-csl}, we study the linear stability and causality of the third-order viscous hydrodynamics. Finally, we summarize and conclude our work in Sec.~\eqref{smry}. Throughout the text, we use natural units where $ \hbar = c = k_B =1$. We consider Minkowski metric in this work which is denoted by $g^{\mu\nu}=\text{diag}\left(1,-1,-1,-1\right)$. We use bold font to denote three-vectors and employ center-dot to denote scalar products of both three- and four-vectors, i.e., $a \cdot b = a^0 b^0 - {\boldsymbol a} \cdot {\boldsymbol b}$.


\section{Third-order viscous evolution equation}
\label{3rd-ord}

The hydrodynamic evolution of a system, without net conserved charges, is determined by the conservation equations of energy and momentum. The conserved energy-momentum tensor can be represented in terms of the single-particle phase-space distribution function and decomposed into hydrodynamical tensor degrees of freedom. In this study, we further consider a system of massless particles, resulting in the absence of bulk viscosity. The energy-momentum tensor for such a system can be written as
\begin{equation}\label{energy}
	T^{\mu\nu}=\int dP ~p^{\mu}p^{\nu}f(x,p)=\epsilon u^{\mu}u^{\nu}-P\Delta^{\mu\nu}+\pi^{\mu\nu},
\end{equation}
where, $dP \equiv g\,d^3{{\boldsymbol p}}/\left[(2\pi)^3|{\boldsymbol p}|\right]$ is Lorentz invariant momentum integral measure with $g$ being the degeneracy factor. Here, $p^{\mu}$ is the particle four-momentum and $f(x,p)$ is the single-particle phase-space distribution function with $x^\mu$ representing the position four-vector. In the tensor decomposition $\epsilon$ and $P$ are energy density and thermodynamic pressure, respectively, $\pi^{\mu\nu}$ is the shear-stress tensor and $u^{\mu}$ is the fluid four-velocity defined in the Landau frame, $u_{\mu}T^{\mu\nu}=\epsilon u^{\nu}$. Moreover, we have the orthogonality condition $u_\mu\pi^{\mu\nu}=0$ and introduced the notation $\Delta^{\mu\nu} \equiv g^{\mu\nu} - u^{\mu}u^{\nu}$ as the projection operator orthogonal to $u^{\mu}$. 

The hydrodynamic equations for evolution of $\epsilon$ and $u^{\mu}$ are obtained from the energy-momentum conservation equations, $\partial_\mu T^{\mu\nu}=0$, and can be written as
\begin{align}
	\dot{\epsilon}+\left(\epsilon+P\right)\theta - \pi^{\mu\nu}\sigma_{\mu\nu} &= 0, \label{evolution1} \\
	\left(\epsilon+P\right)\dot{u}^{\alpha}-\nabla^{\alpha}P+\Delta^{\alpha}_{\nu}\partial_{\mu}\pi^{\mu\nu} &= 0.\label{evolution2}
\end{align}
Here, we have used the notations $\dot{A}\equiv u^{\mu}\partial_{\mu}A$ for the comoving derivative, $\nabla^{\alpha}\equiv \Delta^{\mu\alpha}\partial_{\mu}$ for space-like derivative, $\theta\equiv \partial_{\mu}u^{\mu}$ for expansion scalar and $\sigma^{\mu\nu}\equiv \frac{1}{2}\left(\nabla^{\mu}u^{\nu}+\nabla^{\nu}u^{\mu}\right)-\frac{1}{3}\theta\Delta^{\mu\nu} $ for the velocity stress tensor. Using the Landau matching condition, $\epsilon=\epsilon_{\rm eq}$ with $\epsilon_{\rm eq}$ being the equilibrium energy density, we obtain $\epsilon=3P\propto \beta^{-4}$ in the case of massless particles. The derivative of inverse temperature, $\beta\equiv 1/T$, can be obtained from
Eqs.~\eqref{evolution1} and \eqref{evolution2} as
\begin{align}
	\dot{\beta}=\frac{\beta}{3}\theta-\frac{\beta}{12 P}\pi^{\mu\nu}\sigma_{\mu\nu}, \label{dbeta1} \\
	\nabla^{\alpha}\beta=-\beta\dot{u}^{\alpha}-\frac{\beta}{4P}\Delta^{\alpha}_{\nu}\partial_{\mu}\pi^{\mu\nu}. \label{dbeta2}
\end{align}
These expressions for derivatives of $\beta$ will be further used for obtaining the third-order viscous evolution equation.

In this work, we consider a system of relativistic particles with vanishing chemical potential, close to the local thermodynamic equilibrium. In this case, the single particle phase-space distribution function can be written as $f=f_{\rm eq}+\delta f$, where the deviation from equilibrium is assumed to be small ($\delta f<< f_{\rm eq}$). In this work, we further assume that the equilibrium is described by classical Maxwell-Boltzmann distribution in J\"uttner form, $f_{\rm eq}=\text{exp}\left(-\beta~ u\cdot p\right)$, with $u\cdot p \equiv u_{\mu}p^{\mu}$. Using Eq. \eqref{energy}, the form of $\pi^{\mu\nu}$ can be expressed in terms of $\delta f$ as
\begin{equation}\label{shear}
    \pi^{\mu\nu} = \Delta^{\mu\nu}_{\alpha\beta}\int dP \, p^{\alpha} \, p^{\beta} \, \delta f,
\end{equation}
where, $\Delta^{\mu\nu}_{\alpha\beta}\equiv \frac{1}{2}\!\left( \Delta^{\mu}_{\alpha}\Delta^{\nu}_{\beta} + \Delta^{\mu}_{\beta}\Delta^{\nu}_{\alpha} \right) - \frac{1}{3}\Delta^{\mu\nu}\Delta_{\alpha\beta}$ is a traceless and doubly symmetric projection operator, which is orthogonal to $u_{\mu}$ as well as $\Delta_{\mu\nu}$.
   
The non-equilibrium phase-space distribution function can be obtained by solving the kinetic equation such as Boltzmann equation. The relativistic Boltzmann transport equation under relaxation-time approximation (RTA) for the collision term is given by \cite{Anderson:1974nyl}
\begin{equation}\label{boltz}
    p^{\mu}\partial_{\mu}f = - \frac{u\cdot p}{\tau_R}\,\delta f,
\end{equation}
where $\tau_R$ is the relaxation time. We note that RTA is a simple yet useful model for the collision kernel, which satisfies fundamental conservation equations when $\tau_R$ is independent of momenta and $u^{\mu}$ is defined in the Landau frame~\cite{Anderson:1974nyl, DeGroot:1980dk}. In order to calculate dissipative corrections to the distribution function, Chapman-Enskog-like iterative solution to the RTA Boltzmann equation is considered, where the particle distribution function is expanded in powers of space-time gradients about its equilibrium value \cite{Chapman:1970}
\begin{equation}\label{CE-like-exp}
    f = f_{\rm eq}+\delta f, \quad \delta f = \delta f^{(1)} + \delta f^{(2)} + \cdots.
\end{equation}
Here, $\delta f^{(n)}$ is the non-equilibrium correction which is $n$-th order in derivatives. 

For first- and second-order in derivatives, we have,
\begin{align}
\delta f^{(1)} & =-\frac{\tau_R}{u \cdot p} p^\mu \partial_\mu f_{\rm eq}, \label{delf1} \\
\delta f^{(2)} & =\frac{\tau_R}{u \cdot p} p^\mu p^\nu \partial_\mu\left(\frac{\tau_R}{u \cdot p} \partial_\nu f_{\rm eq}\right) \label{delf2}.
\end{align}
The first-order expression of $\pi^{\mu\nu}$, which is the relativistic version of the Navier-Stokes equation, can be calculated by using $\delta f = \delta f^{(1)}$ from Eq.~\eqref{delf1} in Eq.~\eqref{shear} and is obtained as 
\begin{equation}\label{pimunu1}
    \pi^{\mu\nu} = 2 \tau_R\beta_{\pi}\sigma^{\mu\nu}, \quad \beta_{\pi}=\frac{4}{5}P.
\end{equation}
Using the above expression in Eq.~\eqref{delf1}, the first order viscous correction to the distribution function is obtained as
\begin{equation}\label{deltaf1}
    \delta f_1 = \frac{\beta f_{\rm eq}}{2 \beta_\pi(u \cdot p)} p^\alpha p^\beta \pi_{\alpha \beta} + \mathcal{O}\left(\partial^2\right).
\end{equation}
The above expression was shown to have several desirable features in the context of particle production in heavy-ion collision~\cite{Bhalerao:2013pza}.

Following the methodology discussed in Ref.~\cite{Denicol:2010xn}, the evolution equation for shear stress tensor can be obtained by taking the comoving derivative of Eq.~\eqref{shear} as
\begin{equation}\label{pi_dot}
    \dot{\pi}^{<\mu\nu>} = \Delta^{\mu\nu}_{\alpha\beta}\int dP ~p^{\alpha}~p^{\beta}~\delta\dot{f}.
\end{equation}
In the above equation, $\delta\dot{f}$ can be obtained by rewriting Eq. \eqref{boltz} as
\begin{equation}\label{deltafdot}
	\delta\dot{f}=-\dot{f}_{\rm eq}-\frac{1}{u.p}p^{\mu}\nabla_{\mu}f-\frac{\delta f}{\tau_R}.
\end{equation}
Using the above expression of $\delta\dot{f}$, Eq. \eqref{pi_dot} becomes,
\begin{equation}\label{pidot}
\dot{\pi}^{\langle\mu\nu\rangle}+\frac{\pi^{\mu\nu}}{\tau_R}=-\Delta_{\alpha \beta}^{\mu\nu} \int d p p^\alpha p^\beta\left(\dot{f}_{\rm eq}+\frac{1}{u \cdot p} p^\gamma \nabla_\gamma f\right) .
\end{equation}
It is important to observe that in order to obtain second-order shear evolution equation, the distribution function in Eq.~\eqref{pidot} needs to be computed up to first order, i.e., $\delta f_1$. Therefore, second order shear-evolution equation is obtained by substituting Eq.~\eqref{deltaf1} in Eq.~\eqref{pidot}, as~\cite{Jaiswal:2013npa}
\begin{align}\label{pidot-2nd}
\dot{\pi}^{\langle\mu\nu\rangle}+\frac{\pi^{\mu\nu}}{\tau_\pi}=  2 \beta_\pi \sigma^{\mu\nu}+2 \pi_\gamma^{\langle\mu} \omega^{\nu\rangle \gamma} 
-\frac{10}{7} \pi_\gamma^{\langle\mu} \sigma^{\nu\rangle \gamma}-\frac{4}{3} \pi^{\mu\nu} \theta,
\end{align}
where $\omega^{\mu\nu}\equiv \left(\nabla^{\mu}u^{\nu}-\nabla^{\nu}u^{\mu}\right)/2$ is the vorticity tensor. Here, $\tau_{\pi}=\tau_R$ is the shear relaxation time, which is obtained to be as $\tau_{\pi}=\eta/\beta_{\pi}$ by comparing the first-order evolution Eq.~\eqref{pimunu1} with the relativistic Navier-Stokes equation $\pi^{\mu\nu}=2\,\eta\,\sigma^{\mu\nu}$, with $\eta$ being the coefficient of shear viscosity. 

For the third-order shear evolution equation, the distribution function on the right-hand side of Eq.~\eqref{pidot} needs to be computed up to  second order in viscous corrections, i.e., $\delta f = \delta f_1 + \delta f_2$. Using the derivatives of $\beta$ from Eqs.~\eqref{dbeta1} and \eqref{dbeta2}, as well as Eq. \eqref{pidot-2nd} for $\sigma^{\mu\nu}$ in Eqs. \eqref{delf1} and \eqref{delf2}, second-order viscous correction to the distribution function is obtained as~\cite{Bhalerao:2013pza}
\begin{align}
   	\delta f_2 = \frac{f_{\rm eq} \beta}{\beta_\pi} &\left[ \frac{5}{14 \beta_\pi(u \cdot p)} p^\alpha p^\beta \pi_\alpha^\gamma \pi_{\beta \gamma} - \frac{\tau_\pi}{u \cdot p} p^\alpha p^\beta \pi_\alpha^\gamma \omega_{\beta \gamma}\right.
   	 - \frac{\tau_\pi}{3(u \cdot p)} p^\alpha p^\beta \pi_{\alpha \beta} \theta  + \frac{6 \tau_\pi}{5} p^\alpha \dot{u}^\beta \pi_{\alpha \beta} \nonumber\\
     &- \frac{(u \cdot p)}{70 \beta_\pi} \pi^{\alpha \beta} \pi_{\alpha \beta} - \frac{\tau_\pi}{5} p^\alpha\left(\nabla^\beta \pi_{\alpha \beta}\right) + \frac{3 \tau_\pi}{(u \cdot p)^2}~ p^\alpha p^\beta p^\gamma~ \pi_{\alpha \beta}~ \dot{u}_\gamma
   	  - \frac{\tau_\pi}{2(u \cdot p)^2} p^\alpha p^\beta p^\gamma\left(\nabla_\gamma \pi_{\alpha \beta}\right) \nonumber\\
   	& \left. + \frac{\beta+(u \cdot p)^{-1}}{4(u \cdot p)^2 \beta_\pi}\left(p^\alpha p^\beta \pi_{\alpha \beta}\right)^2\right] + \mathcal{O}\left(\partial^3\right). \label{deltaf2}
\end{align}
Finally, substituting $\delta f = \delta f_1 + \delta f_2$ from Eqs.~\eqref{deltaf1} and \eqref{deltaf2}, and keeping terms up to cubic order in derivatives, third-order evolution equation is obtained as \cite{Jaiswal:2013vta}
\begin{align}\label{pidot-3rd}
\dot{\pi}^{\langle\mu\nu\rangle}= & -\frac{\pi^{\mu\nu}}{\tau_\pi}+2 \beta_\pi \sigma^{\mu\nu}+2 \pi_\gamma^{\langle\mu} \omega^{\nu\rangle \gamma}-\frac{10}{7} \pi_\gamma^{\langle\mu} \sigma^{\nu\rangle \gamma} 
-\frac{4}{3} \pi^{\mu\nu} \theta+\frac{25}{7 \beta_\pi} \pi^{\rho\langle\mu} \omega^{\nu\rangle \gamma} \pi_{\rho \gamma}-\frac{1}{3 \beta_\pi} \pi_\gamma^{\langle\mu} \pi^{\nu\rangle \gamma} \theta \nonumber\\
& -\frac{38}{245 \beta_\pi} \pi^{\mu\nu} \pi^{\rho \gamma} \sigma_{\rho \gamma}-\frac{22}{49 \beta_\pi} \pi^{\rho\langle\mu} \pi^{\nu\rangle \gamma} \sigma_{\rho \gamma} 
 -\frac{24}{35} \nabla^{\langle\mu}\left(\pi^{\nu\rangle \gamma} \dot{u}_\gamma \tau_\pi\right)+\frac{4}{35} \nabla^{\langle\mu}\left(\tau_\pi \nabla_\gamma \pi^{\nu\rangle \gamma}\right) \nonumber\\
& -\frac{2}{7} \nabla_\gamma\left(\tau_\pi \nabla^{\langle\mu} \pi^{\nu\rangle \gamma}\right)+\frac{12}{7} \nabla_\gamma\left(\tau_\pi \dot{u}^{\langle\mu} \pi^{\nu\rangle \gamma}\right) 
 -\frac{1}{7} \nabla_\gamma\left(\tau_\pi \nabla^\gamma \pi^{\langle\mu\nu\rangle}\right)+\frac{6}{7} \nabla_\gamma\left(\tau_\pi \dot{u}^\gamma \pi^{\langle\mu\nu\rangle}\right) \nonumber\\
& -\frac{2}{7} \tau_\pi \omega^{\rho\langle\mu} \omega^{\nu\rangle \gamma} \pi_{\rho \gamma}-\frac{2}{7} \tau_\pi \pi^{\rho\langle\mu} \omega^{\nu\rangle \gamma} \omega_{\rho \gamma} 
 -\frac{10}{63} \tau_\pi \pi^{\mu\nu} \theta^2+\frac{26}{21} \tau_\pi \pi_\gamma^{\langle\mu} \omega^{\nu\rangle \gamma} \theta.
\end{align}
Linear stability and causality analysis of the above equation was discussed in Ref.~\cite{Brito:2021iqr} and it was found that terms containing second-order space-like derivatives of the shear-stress tensor leads to the occurrence of additional unstable modes. In the above equation, the terms contributing to instability and acausality are $\nabla^{\langle\mu}\left(\tau_\pi \nabla_\gamma \pi^{\nu\rangle \gamma}\right)$, $\nabla_\gamma\left(\tau_\pi \nabla^{<\mu}\pi^{\nu>\gamma}\right)$ and $\nabla_\gamma\left(\tau_\pi \nabla^\gamma \pi^{\langle\mu\nu\rangle}\right)$. In the following, we restore causality in the formulation of relativistic third-order viscous hydrodynamics from RTA kinetic theory using Chapman-Enskog-like iterative solution.


\section{Restoring causality at third-order}
\label{3rd-csl}

In Ref.~\cite{Brito:2021iqr}, authors proposed a mechanism to restore causality in third-order hydrodynamic theories by promoting the space-like gradients of the shear-stress tensor to a new hydrodynamical variable which is a third rank tensor, i.e, 
\begin{equation}\label{eq:csl_mec}
    \nabla^{\langle\mu}\pi^{\nu\lambda\rangle} \to \rho^{\mu\nu\lambda}.
\end{equation}
Here, $A^{\langle\mu\nu\lambda\rangle}\equiv\Delta^{\mu\nu\lambda}_{\alpha\beta\rho} A^{\alpha\beta\rho}$ with a symmetric traceless six rank projection operator, orthogonal to the fluid four-velocity, defined as
\begin{align}
	\Delta_{\alpha \beta \rho}^{\mu\nu \lambda} \,\equiv\, & \frac{1}{6}\left[\Delta_\alpha^\mu\left(\Delta_\beta^\nu \Delta_\rho^\lambda+\Delta_\rho^\nu \Delta_\beta^\lambda\right)+\Delta_\beta^\mu\left(\Delta_\alpha^\nu \Delta_\rho^\lambda+\Delta_\rho^\nu \Delta_\alpha^\lambda\right)+\Delta_\rho^\mu\left(\Delta_\alpha^\nu \Delta_\beta^\lambda+\Delta_\beta^\nu \Delta_\alpha^\lambda\right)\right] \nonumber\\
	& -\frac{1}{15}\left[\Delta^{\mu\nu}\left(\Delta_\alpha^\lambda \Delta_{\beta \rho}+\Delta_\beta^\lambda \Delta_{\alpha \rho}+\Delta_\rho^\lambda \Delta_{\alpha \beta}\right)+\Delta^{\mu \lambda}\left(\Delta_\alpha^\nu \Delta_{\beta \rho}+\Delta_\beta^\nu \Delta_{\alpha \rho}+\Delta_\rho^\nu \Delta_{\alpha \beta}\right)\right. \nonumber\\
	& \left.+\Delta^{\nu \lambda}\left(\Delta_\alpha^\mu \Delta_{\beta \rho}+\Delta_\beta^\mu \Delta_{\alpha \rho}+\Delta_\rho^\mu \Delta_{\alpha \beta}\right)\right] . \label{proj6}
\end{align}
The expression for $\rho^{\mu\nu\lambda}$ in terms of $\delta f$ is given by
\begin{equation}\label{def}
\rho^{\mu\nu\lambda}=\Delta^{\mu\nu\lambda}_{\alpha\beta\rho}\int dP ~\frac{p^{\alpha}p^{\beta}p^{\rho}}{u\cdot p}~\delta f.
\end{equation}
We observe that the above expression for $\rho^{\mu\nu\lambda}$ has the same dimension as that of $\pi^{\mu\nu}$ and corresponds to $r=-1$ for expression of $\rho^{\mu\nu\lambda}_r$, provided in Ref.~\cite{deBrito:2023tgb}.  

We note that the first-order viscous correction to the distribution function does not contribute to the expression of $\rho^{\mu\nu\lambda}$ as substituting Eq.~\eqref{deltaf1} in Eq.~\eqref{def} leads to vanishing result. Using the second-order viscous correction to the distribution function from Eq.~\eqref{deltaf2} in Eq.~\eqref{def}, the ``Navier-Stokes equivalent" result for $\rho^{\mu\nu\lambda}$ is obtained as
\begin{equation}\label{first_order_rho}
    \rho^{\mu\nu\lambda} = \frac{3}{7}\tau_{\pi}\nabla^{<\mu}\pi^{\nu\lambda>}-\frac{18}{7}\tau_{\pi}\dot{u}^{<\mu}\pi^{\nu\lambda>}.
\end{equation}
The above equation states that the expression for $\rho^{\mu\nu\lambda}$ is at-least second-order. By comparing the Eq. \eqref{first_order_rho} with the relativistic ``Navier-Stokes" analog equation, $\rho^{\mu\nu\lambda}=\frac{3}{7}\eta_{\rho}\nabla^{<\mu}\pi^{\nu\lambda>}$, we obtain $\eta_{\rho}=\tau_{\pi}$. In order to compare our result with that obtained using moment method, we calculate the expression for $\rho^{\mu\nu\lambda}_{-1}$ of Ref.~\cite{deBrito:2023tgb} in the massless case
\begin{equation}\label{first_order_rho_mm}
    \rho^{\mu\nu\lambda}_{-1} = \frac{18}{49}\tau_{\pi}\nabla^{<\mu}\pi^{\nu\lambda>}-\frac{18}{7}\tau_{\pi}\dot{u}^{<\mu}\pi^{\nu\lambda>}.
\end{equation}
Comparing Eqs.~\eqref{first_order_rho} and \eqref{first_order_rho_mm}, we observe that the expression for $\eta_\rho$ differs in the two cases. From the iterative Chapman-Enskog-like approach, we obtain $\eta_\rho=\tau_\pi$ whereas the moment method leads to $\eta_\rho=(6/7)\tau_\pi$. 

Up to third-order the terms containing the spatial derivative of $\pi^{\mu\nu}$ in Eq. \eqref{pidot-3rd} can be replaced using Eq.~\eqref{first_order_rho} in terms of derivative of $\rho^{\mu\nu\lambda}$. Therefore, the expression for evolution of shear stress tensor in Eq.~\eqref{pidot-3rd} can be rewritten as
\begin{align}
\dot{\pi}^{\langle\mu\nu\rangle}= & -\frac{\pi^{\mu\nu}}{\tau_\pi}+2 \beta_\pi \sigma^{\mu\nu}+2 \pi_\gamma^{\langle\mu} \omega^{v\rangle \gamma}-\frac{10}{7} \pi_\gamma^{\langle\mu} \sigma^{v\rangle \gamma} 
 -\frac{4}{3} \pi^{\mu\nu} \theta+\frac{24}{7 \beta_\pi} \pi^{\rho\langle\mu} \omega^{v\rangle \gamma} \pi_{\rho \gamma}-\frac{5}{21 \beta_\pi} \pi_\gamma^{\langle\mu} \pi^{v\rangle \gamma} \theta \nonumber \\
& -\frac{52}{245 \beta_\pi} \pi^{\mu\nu} \pi^{\rho \gamma} \sigma_{\rho \gamma}-\frac{15}{49 \beta_\pi} \pi^{\rho\langle\mu} \pi^{v\rangle \gamma} \sigma_{\rho \gamma}-\frac{2}{7} \tau_\pi \omega^{\rho\langle\mu} \omega^{v\rangle \gamma} \pi_{\rho \gamma}-\frac{4}{7} \tau_\pi \pi^{\rho\langle\mu} \omega^{v\rangle \gamma} \omega_{\rho \gamma} 
 -\frac{8}{63} \tau_\pi \pi^{\mu\nu} \theta^2 \nonumber\\
 &+\frac{26}{21} \tau_\pi \pi_\gamma^{\langle\mu} \omega^{v\rangle \gamma} \theta-\nabla_{\gamma}\rho^{\gamma<\mu\nu>}+\frac{1}{7\beta_{\pi}}\pi^{\gamma<\mu}\pi^{\nu>\beta}\omega_{\gamma\beta}. \label{third_order_pi}
\end{align}
The detailed steps for the derivation are provided in Appendix~\eqref{pi_rho}. We immediately observe that the shear stress tensor is now coupled to a novel degree of freedom $\rho^{\mu\nu\lambda}$, which satisfies its own equation of motion. To obtain a complete third-order formulation, it is necessary to derive the evolution equation for $\rho^{\mu\nu\lambda}$ up to third-order.

The evolution equation of $\rho^{\mu\nu\lambda}$ is obtained by taking the comoving derivative of Eq. \eqref{def} and projecting the completely symmetric and traceless part,
\begin{align}
	&\dot{\rho}^{<\mu\nu\lambda>}=\Delta^{\mu\nu\lambda}_{\gamma\delta\sigma}\left[\dot{\Delta}_{\alpha\beta\rho}^{\gamma\delta\sigma}\int dP\frac{p^{\alpha}p^{\beta}p^{\rho}}{(u\cdot p)}\delta f\right]+\Delta_{\alpha\beta\rho}^{\mu\nu\lambda}\int dP~ p^{\alpha}p^{\beta}p^{\rho}\left[\frac{\delta \dot{f}}{(u\cdot p)}-\frac{\delta f}{(u\cdot p)^2}D(u\cdot p)\right],\label{evol}
\end{align}
where, we remind that the notation $\rho^{<\mu\nu\lambda>}\equiv \Delta^{\mu\nu\lambda}_{\gamma\delta\sigma}~\rho^{\gamma\delta\sigma}$ represents the traceless symmetric projection orthogonal to $u^{\mu}$.
Using Eq.~\eqref{deltafdot} in \eqref{evol}, we obtain
\begin{align}\label{eom}
	\dot{\rho}^{<\mu\nu\lambda>}+\frac{1}{\tau_R}\rho^{\mu\nu\lambda} = \Delta^{\mu\nu\lambda}_{\gamma\delta\sigma} \left[ \dot{\Delta}_{\alpha\beta\rho}^{\gamma\delta\sigma} \!\int\! dP\, \frac{p^{\alpha}p^{\beta}p^{\rho}}{(u\cdot p)}\delta f \right] -\Delta_{\alpha\beta\rho}^{\mu\nu\lambda} \!\int\! dP\, p^{\alpha}p^{\beta}p^{\rho}\Bigg[\frac{\dot{f_{\rm eq}}}{(u\!\cdot\! p)}
	+\frac{p^{\gamma}\nabla_{\gamma}f}{(u\!\cdot\! p)^2} +\frac{\delta f}{(u\!\cdot\! p)^2}D(u\!\cdot\! p)\Bigg].
\end{align}
Using the expressions for derivatives of $f_{\rm eq}$ in the above equation, along with Eqs.~\eqref{dbeta1} and \eqref{dbeta2} for derivatives of $\beta$, we obtain
\begin{align}\label{eom1}
\dot{\rho}^{<\mu\nu\lambda>}+\frac{1}{\tau_R}\rho^{\mu\nu\lambda} =&\, \Delta^{\mu\nu\lambda}_{\gamma\delta\sigma}\left[ \dot{\Delta}_{\alpha\beta\rho}^{\gamma\delta\sigma} \!\int\! dP\frac{p^{\alpha}p^{\beta}p^{\rho}}{(u\!\cdot\! p)}\delta f \right]-\Delta^{\mu\nu\lambda}_{\alpha\beta\rho} \!\int\! dP \frac{p^{\alpha}p^{\beta}p^{\rho}}{u\cdot p}\left[\left(u\!\cdot\! p\right)\!\left(\frac{\beta\theta}{3}-\frac{\beta}{12P}\pi^{\gamma\delta}\sigma_{\gamma\delta}\!\right)+\beta p^{\gamma}\dot{u}_{\gamma}\right]f_{\rm eq} \nonumber\\
&-\Delta_{\alpha\beta\rho}^{\mu\nu\lambda}\int dP~p^{\alpha}p^{\beta}p^{\rho}p^{\gamma} \left[\nabla_{\gamma}\!\left(\!\frac{f_{\rm eq}}{(u\!\cdot\! p)^2}\!\right)+\frac{2p^{\sigma}\left(\nabla_{\gamma}u_{\sigma}\right)}{(u\cdot p)^3}f_{\rm eq}+\nabla_{\gamma}\!\left(\!\frac{\delta f}{(u\!\cdot\! p)^2}\!\right) +\frac{2p^{\sigma}\left(\nabla_{\gamma}u_{\sigma}\right)}{(u\cdot p)^3}\delta f \right] \nonumber\\
&-\Delta_{\alpha\beta\rho}^{\mu\nu\lambda}~\int dP~ \frac{p^{\alpha}p^{\beta}p^{\rho}p^{\gamma}\dot{u}_{\gamma}}{(u\cdot p)^2}\delta f.
\end{align}
It is apparent from the form of the above equation that the relaxation time $\tau_R$ can be identified with novel relaxation time $\tau_{\rho}$. 

The six rank tensor, $\Delta_{\mu\nu\lambda}^{\gamma\delta\theta}$ is orthogonal to 4-velocity and second rank tensor, i.e. $\Delta_{\mu\nu\lambda}^{\gamma\delta\theta}u^{\mu}=0$, $\Delta_{\mu\nu\lambda}^{\gamma\delta\theta}\Delta^{\mu\nu}=0$. Using these properties, and Eqs.~\eqref{deltaf1} and \eqref{deltaf2} for $\delta f = \delta f_1 + \delta f_2$, as well as Eqs.~\eqref{dbeta1}, \eqref{dbeta2} for derivatives of $\beta$, the first term on the right-hand-side of Eq.~\eqref{eom1} is obtained in terms of thermodynamic integrals as
\begin{align}
\Delta^{\mu\nu\lambda}_{\gamma\delta\sigma}\left[\dot{\Delta}_{\alpha\beta\rho}^{\gamma\delta\sigma}\int dP\frac{p^{\alpha}p^{\beta}p^{\rho}}{(u\cdot p)}\delta f\right]=&-\frac{3\beta}{\beta_{\pi}}I_{52}^{(2)}\left(\dot{u}^{<\mu}~\pi^{\nu\lambda>}\right)-\frac{6\tau_{\pi}\beta}{\beta_{\pi}}I_{52}^{(2)}~\left(\omega^{\gamma<\mu}\pi^{\nu}_{\gamma}\dot{u}^{\lambda>}\right)\nonumber\\
&-\frac{15\beta}{7\beta_{\pi}^2}I_{52}^{(2)}~\left(\pi^{\gamma<\mu}\pi^{\nu}_{\gamma}\dot{u}^{\lambda>}\right)+\frac{2\beta\tau_{\pi}}{\beta_{\pi}}I_{52}^{(2)}~\theta~\left(\pi^{<\mu\nu}\dot{u}^{\lambda>}\right).
\end{align}	
Now, considering the second term of Eq.~\eqref{eom1} and using properties, $u^{\alpha}\Delta_{\alpha\beta\rho}^{\mu\nu\lambda}=u^{\beta}\Delta_{\alpha\beta\rho}^{\mu\nu\lambda}=u^{\rho}\Delta_{\alpha\beta\rho}^{\mu\nu\lambda}=\Delta^{\alpha\beta}\Delta_{\alpha\beta\rho}^{\mu\nu\lambda}=\Delta^{\alpha\rho}\Delta_{\alpha\beta\rho}^{\mu\nu\lambda}=\Delta^{\rho\beta}\Delta_{\alpha\beta\rho}^{\mu\nu\lambda}=0$, the contribution from second term vanishes. Similarly, the third and fourth terms of Eq. \eqref{eom1} can be calculated and details of the calculation can be found in Appendix \eqref{calculation}. Finally, the evolution equation of $\rho^{\mu\nu\lambda}$ is obtained as
%
%
\begin{align}
	\dot{\rho}^{<\mu\nu\lambda>}+\frac{1}{\tau_\rho}\rho^{\mu\nu\lambda}=&~
	\frac{3}{7}\nabla^{<\mu}\pi^{\nu\lambda>}-~\frac{18}{7}\dot{u}^{<\mu}\pi^{\nu\lambda>}-\frac{187}{81}\rho^{\mu\nu\lambda}\theta-\frac{10}{7}\tau_{\pi}\dot{u}^{<\mu}\pi^{\nu\lambda>}\theta-\frac{36}{7}\tau_{\pi}\omega^{\gamma<\mu}\pi^{\nu}_{\gamma}\dot{u}^{\lambda>}\nonumber \\
 &-\frac{389}{441\beta_{\pi}}\pi^{\gamma<\mu}\pi^{\nu}_{\gamma}\dot{u}^{\lambda>}-\frac{343}{105\beta_{\pi}}\dot{u}_{\gamma}\pi^{\gamma<\mu}\pi^{\nu\lambda>}+\frac{18}{7}\tau_{\pi}\dot{u}_{\gamma}\omega^{\gamma<\mu}\pi^{\nu\lambda>}-\frac{6}{7}\tau_{\pi}\omega^{\gamma<\mu}\nabla_{\gamma}\pi^{\nu\lambda>} \nonumber\\
&- \frac{6}{7}\tau_{\pi}\omega^{\gamma<\mu}\nabla^{\nu}\pi^{\lambda>}_{\gamma}-\frac{6}{7}\tau_{\pi}\pi_{\gamma}^{<\mu}\nabla^{\nu}\omega^{\lambda>\gamma}-\frac{47}{63\beta_{\pi}}\pi^{<\mu\nu}\nabla_{\gamma}\pi^{\lambda>\gamma}-\frac{11}{21\beta_{\pi}}\pi^{\gamma<\mu}\nabla_{\gamma}\pi^{\nu\lambda>} \nonumber\\
&-\frac{665}{441\beta_{\pi}}\pi^{\gamma<\mu}\nabla^{\nu}\pi^{\lambda>}_{\gamma}. \label{rho}
\end{align}
By comparing second-order terms in the above equation with the ``Navier-Stokes"-equivalent equation, $\rho^{\mu\nu\lambda}=\frac{3}{7}\eta_{\rho}\nabla^{<\mu}\pi^{\nu\lambda>}$, we obtain $\eta_{\rho}=\tau_{\pi}$. Furthermore, in relaxation-time approximation, all relaxation times are identical, i.e. $\tau_{\pi}=\tau_{\rho}=\tau_R$.

We can compare the  relativistic third-order evolution equation for viscous hydrodynamic derived in this work with that obtained in \cite{deBrito:2023tgb}. In the latter work, the derivation of third-order relativistic dissipative fluid dynamics employed the method of moments, incorporating two novel degrees of freedom. These degrees of freedom were associated with the irreducible tensors of rank 3 and 4. For ease of comparison, we express the third order evolution equation obtained in Ref. \cite{deBrito:2023tgb} as
\begin{equation}\label{dpi_denicol}
\tau_{\pi}\dot{\pi}^{\mu\nu}+\pi^{\mu\nu}=2\eta\sigma^{\mu\nu}+2\tau_{\pi}\pi^{<\mu}_{\lambda}\omega^{\nu>\lambda}-\delta_{\pi\pi}\pi^{\mu\nu}\theta-\tau_{\pi\pi}\pi^{<\mu}_{\lambda}\sigma^{\nu>\lambda}-\tau_{\pi}\gamma_{-1}^{\Omega}\Delta^{\mu\nu}_{\alpha\beta}\nabla_{\lambda}\Omega^{\alpha\beta\lambda}+\tau_{\pi}\tau_{\pi\Omega}\dot{u}_{\alpha}\Omega^{\mu\nu\alpha}-\tau_{\pi}\gamma_{-2}^{\Theta}\Theta^{\mu\nu\alpha\beta}\sigma_{\alpha\beta} .
\end{equation}
Here, $\Omega^{\mu\nu\lambda}$ and $\Theta^{\mu\nu\alpha\beta}$ are the two novel degrees of freedom. They satisfy their own equation of motion and were obtained as
\begin{align}\label{omega}
\tau_{\Omega} \dot{\Omega}^{\langle\mu\nu \alpha\rangle}+\Omega^{\mu\nu \alpha}  =~&\delta_{\Omega \Omega} \Omega^{\mu\nu \alpha} \theta+3 \tau_{\Omega} \Omega^{\lambda\langle\mu\nu} \omega^{\alpha\rangle}{ }_\lambda+\tau_{\Omega \Omega} \sigma_\lambda^{\langle\mu} \Omega^{\nu \alpha\rangle \lambda}+\frac{3}{7} \eta_{\Omega} \nabla^{\langle\mu} \pi^{\nu \alpha\rangle} 
 -3 \tau_{\Omega} \gamma_1^\pi \pi^{\langle\mu\nu} \dot{u}^{\alpha\rangle}\nonumber\\
 &+\tau_{\Omega \Theta} \Theta^{\mu\nu \alpha \beta} \dot{u}_\beta-\tau_{\Omega} \gamma_{-1}^{\Theta} \Delta_{\lambda \sigma \rho}^{\mu\nu \alpha} \nabla_\beta \Theta^{\lambda \sigma \rho \beta} ,
\end{align}
\begin{align}\label{Theta}
\tau_{\Theta} \dot{\Theta}^{\langle\mu\nu \alpha \beta\rangle}+\Theta^{\mu\nu \alpha \beta}  =~&\delta_{\Theta \Theta} \Theta^{\mu\nu \alpha \beta} \theta+4 \tau_{\Theta} \Theta^{\lambda\langle\mu\nu \alpha} \omega^{\beta\rangle}{ }_\lambda+\tau_{\Theta \Theta} \sigma_\lambda^{\langle\mu} \Theta^{\nu \alpha \beta\rangle \lambda}+\ell_{\Theta \pi} \sigma^{\langle\mu\nu} \pi^{\alpha \beta\rangle}+ 
 +\ell_{\Theta \Omega} \nabla^{\langle\mu} \Omega^{\nu \alpha \beta\rangle}\nonumber\\
 &+\tau_{\Theta \Omega} \dot{u}^{\langle\mu} \Omega^{\nu \alpha \beta\rangle}.
\end{align}
Here, third rank tensor $\Omega^{\mu\nu\lambda}$ corresponds to our $\rho^{\mu\nu\lambda}$ via the relation $\rho^{\mu\nu\lambda}_{-1}=\gamma^\Omega_{-1}\Omega^{\mu\nu\lambda}$. The value of different coefficients in Eqs. \eqref{dpi_denicol} to \eqref{Theta} for the classical and massless case can be found in Ref. \cite{deBrito:2023tgb} and some of them are given below:
\begin{equation}
\begin{aligned}
& \gamma^\Omega_{-1} =\frac{\beta}{7} , ~ \gamma^\Theta_{-2} = \frac{\beta^2}{72}, ~ \frac{\delta_{\pi\pi}}{\tau_{\pi}}=\frac{4}{3}, ~ \frac{\tau_{\pi\pi}}{\tau_{\pi}}=\frac{10}{7}, ~ \tau_{\pi\Omega} = \frac{\beta}{7}, ~ \delta_{\Omega \Omega}=-\frac{5}{3} \tau_{\Omega}, ~ \tau_{\Omega \Omega}=-\frac{7}{3} \tau_{\Omega}, ~ \eta_{\Omega}=\frac{6}{\beta} \tau_{\Omega}.
\end{aligned}
\end{equation}
The viscous coefficient associated with new dynamical degree of freedom, the third rank tensor, is found to be $\eta_{\rho}=\tau_{\pi}$ in our case and $\eta_{\Omega}=\frac{6}{\beta}\tau_{\Omega}$ in \cite{deBrito:2023tgb}. We note that the equation of motion for $\rho^{\mu\nu\lambda}$ in our case, given in Eq. \eqref{rho}, is not coupled to a new degree of freedom, hence it forms a closed set of equations. Therefore, it is not required to find a evolution equation for a $4$-th rank tensor in the present case. 


\section{Linear stability and causality analysis}
\label{stb-csl}

In this section, we analyze the stability of the third-order fluid dynamical formulation in its linear regime, which is done via a linear stability analysis. In this analysis, the system is assumed to be initially in a global equilibrium state and performs small perturbations on the hydrodynamic variables around such state. In a stable fluid-dynamical formulation, the perturbations are damped with time, allowing the fluid to return to its initial global equilibrium state. In contrast, an unstable theory would cause the perturbations to increase exponentially over time and the system will never return to its initial equilibrium state. In the present work, the linear contributing terms in the evolution equation of $\pi^{\mu\nu}$ in Eq.~\eqref{third_order_pi} and $\rho^{\mu\nu\lambda}$ in Eq.~\eqref{rho} are
\begin{equation}\label{pi_linear}
    \dot{\pi}^{<\mu\nu>}= -\frac{\pi^{\mu\nu}}{\tau_{\pi}}+2\beta_{\pi}\
    \sigma^{\mu\nu}-\nabla_{\gamma}\rho^{\gamma<\mu\nu>}+\cdots
\end{equation}
and 
\begin{equation}\label{rho_linear}
    \dot{\rho}^{<\mu\nu\lambda>}=\frac{1}{\tau_\rho}\rho^{\mu\nu\lambda}+\frac{3}{7}\nabla^{<\mu}\pi^{\nu\lambda>} +\cdots ,
\end{equation}
respectively. The dots denote nonlinear terms, and will not contribute to the linear stability analysis. 

We consider the following perturbations on the hydrodynamic variables
\begin{equation}\label{perturb}
    \epsilon = \epsilon_0+\delta\epsilon,\quad u^{\mu}=u_{0}^{\mu}+\delta u^{\mu},\quad \pi^{\mu\nu}=\delta \pi^{\mu\nu},\quad \rho^{\mu\nu\lambda} = \delta\rho^{\mu\nu\lambda},
\end{equation}
with $\epsilon_0$ and $u^{\mu}_{0}$ being the energy-density and 4-velocity of the background global equilibrium, while $\pi^{\mu\nu}$ and $\rho^{\mu\nu\lambda}$ are the perturbations themselves. Under these perturbations, the linearized Eqs. \eqref{pi_linear} and \eqref{rho_linear} becomes
\begin{align}
\tau_\pi D_0 \delta \pi^{\mu\nu}+\delta \pi^{\mu\nu}= & \eta\left(\nabla_0^\mu \delta u^\nu+\nabla_0^\nu \delta u^\mu-\frac{2}{3} \Delta_0^{\mu\nu} \partial_\lambda \delta u^\lambda\right) 
-\frac{3}{7}\tau_\pi \nabla_\gamma^0 \delta \rho^{\gamma \mu\nu},
\\
\tau_\rho D_0 \delta \rho^{\mu\nu\lambda}+\delta \rho^{\mu\nu\lambda}= & \eta_\rho\Bigg[\frac{1}{7}\left(\nabla_0^\lambda \delta \pi^{\mu\nu}+\nabla_0^\nu\delta \pi^{\mu\lambda}+\nabla_0^\mu\delta \pi^{\nu\lambda}\right)
-\frac{2}{35}\left(\Delta_0^{\mu\nu} \nabla_\alpha^0 \delta \pi^{\lambda \alpha}+\Delta_0^{\mu \lambda} \nabla_\alpha^0 \delta \pi^{\nu \alpha}\right) \nonumber\\
& +\left.\Delta_0^{\nu\lambda} \nabla_\alpha^0 \delta \pi^{\mu \alpha}\right)\Bigg] .
\end{align}
 The causality and stability of the above linear equations were studied in \cite{Brito:2021iqr} and the following constraints were obtained on transport coefficients,
\begin{equation}\label{causal}
\left[3 \tau_\pi\left(1-c_{\mathrm{s}}^2\right)-4 \frac{\eta}{\epsilon_0+P}\right] \tau_\rho>\frac{27}{35} \eta_\rho \tau_\pi\left(1-c_{\mathrm{s}}^2\right),
\end{equation}
\begin{equation}\label{stable}
3\left(1-\mathrm{c}_{\mathrm{s}}^2\right) \tau_\pi \geq \frac{4 \eta}{\epsilon_0+P} .
\end{equation}
Eq. \eqref{causal} necessitates the existence of non-zero timescale $\tau_{\rho}$, which is essential for both causality and stability. For the case of classical and massless particles, the above conditions simplify to
\begin{align}
    &\eta_{\rho}<\frac{7}{3}\tau_{\rho},\\\nonumber
    &\tau_{\pi}\geq \frac{2\eta}{\epsilon_0+P}.
\end{align}
Our results for the transport coefficient, $\eta_{\rho}=\tau_{\rho}=\tau_{\pi}=5\eta/(\epsilon+P)$, is thus consistent with the conditions listed above. Therefore, we conclude that the present formulation of relativistic third-order viscous hydrodynamics is linearly causal and stable.


\section{Summary and outlook}
\label{smry}

In summary, we have derived the causal and stable third-order viscous hydrodynamics from the relativistic Boltzmann transport equation using the iterative Chapman-Enskog-like expansion. We demonstrated that a causal relativistic third-order theory requires the inclusion of a new dynamical degree of freedom, i.e., an irreducible tensor of rank 3. This feature of our formulation differs from the theory developed using method of moments, which necessitates the inclusion of two irreducible tensors, one of rank 3 and another of rank 4. We derived the equation of motion of new degree of freedom, $\rho^{\mu\nu\lambda}$ and calculated its associated transport coefficients. Additionally, by considering perturbations around a global equilibrium state, we showed that these transport coefficients are in compliance with stability and causality constraints.

Looking forward, it will be interesting to extend the present formalism to a more general case of system of massive particles and as well as to a system of particles with conserved charges. This would require one to calculate the expressions for second-order viscous corrections to the distribution function in these two cases. These involved calculations, within the present framework, is left for future work.


\section{ACKNOWLEDGMENTS}
P. P. is thankful for the kind hospitality of the National Institute of Science Education and Research (NISER) Bhubaneswar, Jatni, where a part of this work was done.


\appendix


\addappheadtotoc

\begin{appendix}
	\renewcommand{\theequation}{A.\arabic{equation}}
 
 \section{ Evolution equation of $\pi^{\mu\nu}$}
 \label{pi_rho}
We have the third-order evolution equation for the shear stress tensor, Eq. \eqref{pidot-3rd} as
\begin{align}
	 \dot{\pi}^{\langle\mu\nu\rangle}=&-\frac{\pi^{\mu\nu}}{\tau_\pi}+2 \beta_\pi \sigma^{\mu\nu}+2 \pi_\gamma^{\langle\mu} \omega^{\nu\rangle \gamma}-\frac{10}{7} \pi_\gamma^{\langle\mu} \sigma^{v\rangle \gamma} 
	 -\frac{4}{3} \pi^{\mu\nu} \theta+\frac{25}{7 \beta_\pi} \pi^{\rho\langle\mu} \omega^{\nu\rangle \gamma} \pi_{\rho \gamma}-\frac{1}{3 \beta_\pi} \pi_\gamma^{\langle\mu} \pi^{\nu\rangle \gamma} \theta \nonumber\\
	& -\frac{38}{245 \beta_\pi} \pi^{\mu\nu} \pi^{\rho \gamma} \sigma_{\rho \gamma}-\frac{22}{49 \beta_\pi} \pi^{\rho\langle\mu} \pi^{\nu\rangle \gamma} \sigma_{\rho \gamma} 
	 \bm{\underbrace{-\frac{24}{35} \nabla^{\langle\mu}\left(\pi^{\nu\rangle \gamma} \dot{u}_\gamma \tau_\pi\right)}_{1}+\underbrace{\frac{4}{35} \nabla^{\langle\mu}\left(\tau_\pi \nabla_\gamma \pi^{\nu\rangle \gamma}\right)}_{2}} \nonumber\\
	& \bm{\underbrace{-\frac{2}{7} \nabla_\gamma\left(\tau_\pi \nabla^{\langle\mu} \pi^{\nu\rangle \gamma}\right)}_{3}+\underbrace{\frac{12}{7} \nabla_\gamma\left(\tau_\pi \dot{u}^{\langle\mu} \pi^{\nu\rangle \gamma}\right)}_{4}} ~~
	\bm{\underbrace{-\frac{1}{7} \nabla_\gamma\left(\tau_\pi \nabla^\gamma \pi^{\langle\mu\nu\rangle}\right)}_{5}+\underbrace{\frac{6}{7} \nabla_\gamma\left(\tau_\pi \dot{u}^\gamma \pi^{\langle\mu\nu\rangle}\right)}_{6}} \nonumber\\
	& -\frac{2}{7} \tau_\pi \omega^{\rho\langle\mu} \omega^{\nu\rangle \gamma} \pi_{\rho \gamma}-\frac{2}{7} \tau_\pi \pi^{\rho(\mu} \omega^{\nu\rangle \gamma} \omega_{\rho \gamma} 
	-\frac{10}{63} \tau_\pi \pi^{\mu\nu} \theta^2+\frac{26}{21} \tau_\pi \pi_\gamma^{\langle\mu} \omega^{\nu\rangle \gamma} \theta . 
\end{align}	
The underbracketed six terms in the above equation contains the second order space-like derivative of shear stress tensor which lead to the additional unstable modes. Writing the above six terms separately as
\begin{align}
    \dot{\pi}^{<\mu\nu>}_{(6)}= &-\frac{24}{35} \nabla^{\langle\mu}\left(\pi^{\nu\rangle \gamma} \dot{u}_\gamma \tau_\pi\right)+\frac{4}{35} \nabla^{\langle\mu}\left(\tau_\pi \nabla_\gamma \pi^{\nu\rangle \gamma}\right) 
	 -\frac{2}{7} \nabla_\gamma\left(\tau_\pi \nabla^{\langle\mu} \pi^{\nu\rangle \gamma}\right)+\frac{12}{7} \nabla_\gamma\left(\tau_\pi \dot{u}^{\langle\mu} \pi^{\nu\rangle \gamma}\right) \nonumber\\
	&-\frac{1}{7} \nabla_\gamma\left(\tau_\pi \nabla^\gamma \pi^{\langle\mu\nu\rangle}\right)+\frac{6}{7} \nabla_\gamma\left(\tau_\pi \dot{u}^\gamma \pi^{\langle\mu\nu\rangle}\right),\\\nonumber
 =&-\frac{24}{35} \nabla^{\langle\mu}\left(\pi^{\nu\rangle \gamma} \dot{u}_\gamma \tau_\pi\right)+\frac{12}{7} \nabla_\gamma\left(\tau_\pi \dot{u}^{\langle\mu} \pi^{\nu\rangle \gamma}\right)+\frac{6}{7} \nabla_\gamma\left(\tau_\pi \dot{u}^\gamma \pi^{\langle\mu\nu\rangle}\right)+\frac{1}{7}\nabla_{\gamma}\Big(\frac{4}{5}g^{\alpha\gamma}\Delta^{\mu\nu}_{\alpha\beta}\tau_{\pi}\nabla_{\lambda}\pi^{\beta\lambda}\\
& -2\Delta^{\mu\nu}_{\alpha\beta}\tau_{\pi}\nabla^{\alpha}\pi^{\beta\gamma}-\Delta^{\mu\nu}_{\alpha\beta}\tau_{\pi}\nabla^{\gamma}\pi^{\alpha\beta}\Big)-\frac{4}{35}\left(\nabla^{\alpha}\Delta^{\mu\nu}_{\alpha\beta}\tau_{\pi}\nabla_{\lambda}\pi^{\beta\lambda}\right)-\frac{1}{7}\nabla_{\gamma}\left(\tau_{\pi}\pi^{\alpha\beta}\nabla^{\gamma}\Delta^{\mu\nu}_{\alpha\beta}\right).\label{a2}
\end{align}
 Using Eq. \eqref{first_order_rho}, we can obtain $\nabla_{\gamma}\rho^{\gamma<\mu\nu>}$ as follows,

\begin{align}
    \nabla_{\gamma}\rho^{\gamma<\mu\nu>}= &\,\nabla_{\gamma}\left[\Delta^{\mu\nu}_{\alpha\beta}\left(\frac{3}{7}\tau_{\pi}\nabla^{<\gamma}\pi^{\alpha\beta>}-\frac{18}{7}\tau_{\pi}\dot{u}^{<\gamma}\pi^{\alpha\beta>}\right)\right],\\
    =&\,\nabla_{\gamma}\left(\frac{3}{7}\tau_{\pi}\Delta^{\mu\nu}_{\alpha\beta}\Delta^{\gamma\alpha\beta}_{\rho\sigma\delta}\nabla^{\rho}\pi^{\sigma\delta}\right)-\frac{18}{7}\nabla_{\gamma}\left(\tau_{\pi}\Delta^{\mu\nu}_{\alpha\beta}\Delta^{\gamma\alpha\beta}_{\rho\sigma\delta}\dot{u}^{\rho}\pi^{\sigma\delta}\right),\\
    =&-\frac{1}{7}\nabla_{\gamma}\Big(\frac{4}{5}\Delta^{\alpha\gamma}\Delta^{\mu\nu}_{\alpha\sigma}\tau_{\pi}\nabla_{\delta}\pi^{\sigma\delta}-2\Delta^{\mu\nu}_{\rho\delta}\tau_{\pi}\nabla^{\rho}\pi^{\gamma\delta}-\Delta^{\mu\nu}_{\sigma\delta}\tau_{\pi}\nabla^{\gamma}\pi^{\sigma\delta}\Big)-\frac{18}{7}\nabla_{\gamma}\left(\tau_{\pi}\Delta^{\mu\nu}_{\alpha\beta}\Delta^{\gamma\alpha\beta}_{\rho\sigma\delta}\dot{u}^{\rho}\pi^{\sigma\delta}\right).\label{a3}
\end{align}

On using Eq. \eqref{a3}, Eq. \eqref{a2} can be written as

\begin{align}\label{a4}
  \dot{\pi}^{<\mu\nu>}_{(6)}= &-\frac{24}{35} \nabla^{\langle\mu}\left(\pi^{\nu\rangle \gamma} \dot{u}_\gamma \tau_\pi\right)  +\frac{12}{7} \nabla_\gamma\left(\tau_\pi \dot{u}^{\langle\mu} \pi^{\nu\rangle \gamma}\right)+\frac{6}{7} \nabla_\gamma\left(\tau_\pi \dot{u}^\gamma \pi^{\langle\mu\nu\rangle}\right)-\nabla_{\gamma}(\Delta^{\mu\nu}_{\alpha\beta}\rho^{\gamma\alpha\beta}) \nonumber\\
  &-\frac{18}{7}\nabla_{\gamma}\left(\tau_{\pi}\Delta^{\mu\nu}_{\alpha\beta}\Delta^{\gamma\alpha\beta}_{\rho\sigma\delta}\dot{u}^{\rho}\pi^{\sigma\delta}\right)-\frac{4}{35}\left(\nabla^{\alpha}\Delta^{\mu\nu}_{\alpha\beta}\tau_{\pi}\nabla_{\lambda}\pi^{\beta\lambda}\right)-\frac{1}{7}\nabla_{\gamma}\left(\tau_{\pi}\pi^{\alpha\beta}\nabla^{\gamma}\Delta^{\mu\nu}_{\alpha\beta}\right).
\end{align}
Above equation can be further simplified by using the following identity,
\begin{equation}
   \frac{18}{7} \nabla_{\gamma}\left(\tau_{\pi}\Delta^{\mu\nu}_{\alpha\beta}\Delta^{\gamma\alpha\beta}_{\rho\sigma\delta}\dot{u}^{\rho}\pi^{\sigma\delta}\right)=\frac{6}{7}\nabla_{\gamma}\left(\dot{u}^{\gamma}\pi^{\mu\nu}\right)+\frac{12}{7}\nabla_{\gamma}\left(\dot{u}^{\rho}\pi^{\gamma\delta}\Delta^{\mu\nu}_{\rho\delta}\right)-\frac{24}{35}\nabla_{\gamma}\left(\dot{u}_{\delta}\Delta^{\gamma\alpha}\Delta^{\mu\nu}_{\alpha\sigma}\pi^{\sigma\delta}\right),
\end{equation}
and hence we obtain Eq. \eqref{a4} as
\begin{align}\label{a6}
    \dot{\pi}^{<\mu\nu>}_{(6)} =& -\nabla_{\gamma}(\Delta^{\mu\nu}_{\alpha\beta}\rho^{\gamma\alpha\beta})-\frac{4}{35}\tau_{\pi}\left(\nabla^{\alpha}\Delta^{\mu\nu}_{\alpha\beta}\right)\nabla_{\lambda}\pi^{\beta\lambda}-\frac{1}{7}\nabla_{\gamma}\left(\tau_{\pi}\pi^{\alpha\beta}\nabla^{\gamma}\Delta^{\mu\nu}_{\alpha\beta}\right)+\frac{24}{35}\tau_{\pi}\pi^{\beta\lambda}\dot{u}_{\lambda}\nabla^{\alpha}\Delta^{\mu\nu}_{\alpha\beta},\\
    =&-\nabla_{\gamma}(\Delta^{\mu\nu}_{\alpha\beta}\rho^{\gamma\alpha\beta})-\frac{1}{7}\left(\nabla_{\gamma}\tau_{\pi}\right)\pi^{\alpha\beta}\nabla^{\gamma}\left(\Delta^{\mu\nu}_{\alpha\beta}\right)-\frac{\tau_{\pi}}{7}\pi^{\alpha\beta}~\nabla^2\left(\Delta^{\mu\nu}_{\alpha\beta}\right)+\frac{24}{35}\tau_{\pi}\pi^{\beta\lambda}\dot{u}_{\lambda}\nabla^{\alpha}\Delta^{\mu\nu}_{\alpha\beta} \nonumber\\
    &-\frac{\tau_{\pi}}{7}\left(\frac{4}{5}g^{\alpha}_{\gamma}\nabla_{\lambda}\pi^{\beta\lambda}+\nabla_{\gamma}\pi^{\alpha\beta}\right)\left(\nabla^{\gamma}\Delta^{\mu\nu}_{\alpha\beta}\right).
\end{align}
Now, we will contract the above equation with $\Delta^{\rho\sigma}_{\mu\nu}$ for symmetric traceless contribution. In process of doing it, we obtain $\Delta^{\rho\sigma}_{\mu\nu}\nabla^{\gamma}\left(\Delta^{\mu\nu}_{\alpha\beta}\right)$ as follows
\begin{equation}
\Delta^{\rho\sigma}_{\mu\nu}\nabla^{\gamma}\left(\Delta^{\mu\nu}_{\alpha\beta}\right)  = \frac{1}{2}\left[-\left(\nabla^{\gamma}u^{\mu}\right)u_{\alpha}\left(\Delta^{\rho\sigma}_{\mu\alpha}\right)-\left(\nabla^{\gamma}u^{\nu}\right)u_{\beta}\left(\Delta^{\rho\sigma}_{\nu\beta}\right)-\left(\nabla^{\gamma}u^{\mu}\right)u_{\beta}\left(\Delta^{\rho\sigma}_{\mu\alpha}\right)-\left(\nabla^{\gamma}u^{\mu}\right)u_{\alpha}\left(\Delta^{\rho\sigma}_{\mu\beta}\right)\right] .
\end{equation}
Similarly, one can calculate $\Delta^{\rho\sigma}_{\mu\nu}~\nabla^{\alpha}\left(\Delta^{\mu\nu}_{\alpha\beta}\right)$ and $\Delta^{\rho\sigma}_{\mu\nu}~\nabla^{2}\left(\Delta^{\mu\nu}_{\alpha\beta}\right)$. Using these derivatives and replacing $u_{\beta}(\nabla_{\lambda}\pi^{\beta\lambda})$ wtih $-\pi^{\beta\lambda}(\nabla_{\lambda}u_{\beta})$
, Eq. \eqref{a6} simplifies to 
\begin{equation}
  \dot{\pi}^{<\rho\sigma>}_{(6)} = -\Delta^{\rho\sigma}_{\mu\nu}(\nabla_{\gamma}\Delta^{\mu\nu}_{\alpha\beta}\rho^{\gamma\alpha\beta})+\frac{1}{7\beta_{\pi}}\pi^{\gamma<\rho}\pi^{\sigma>\alpha}\sigma_{\gamma\alpha}+\frac{2}{21\beta_{\pi}}\theta \pi_{\alpha}^{<\rho}\pi^{\sigma>\alpha}+\frac{2}{63}\tau_{\pi}\theta^2\pi^{\rho\sigma}-\frac{2}{35\beta_{\pi}}\pi^{\rho\sigma}\sigma_{\lambda\beta}\pi^{\beta\lambda} .
\end{equation}
Therefore, we finally obtain the third-order evolution equation of shear stress tensor as
\begin{align}
\dot{\pi}^{\langle\mu\nu\rangle}= & -\frac{\pi^{\mu\nu}}{\tau_\pi}+2 \beta_\pi \sigma^{\mu\nu}+2 \pi_\gamma^{\langle\mu} \omega^{v\rangle \gamma}-\frac{10}{7} \pi_\gamma^{\langle\mu} \sigma^{v\rangle \gamma} 
 -\frac{4}{3} \pi^{\mu\nu} \theta+\frac{24}{7 \beta_\pi} \pi^{\rho\langle\mu} \omega^{v\rangle \gamma} \pi_{\rho \gamma}-\frac{5}{21 \beta_\pi} \pi_\gamma^{\langle\mu} \pi^{v\rangle \gamma} \theta \nonumber\\
& -\frac{52}{245 \beta_\pi} \pi^{\mu\nu} \pi^{\rho \gamma} \sigma_{\rho \gamma}-\frac{15}{49 \beta_\pi} \pi^{\rho\langle\mu} \pi^{v\rangle \gamma} \sigma_{\rho \gamma}-\frac{2}{7} \tau_\pi \omega^{\rho\langle\mu} \omega^{v\rangle \gamma} \pi_{\rho \gamma}-\frac{4}{7} \tau_\pi \pi^{\rho\langle\mu} \omega^{v\rangle \gamma} \omega_{\rho \gamma} 
 -\frac{8}{63} \tau_\pi \pi^{\mu\nu} \theta^2 \nonumber\\
 &+\frac{26}{21} \tau_\pi \pi_\gamma^{\langle\mu} \omega^{v\rangle \gamma} \theta-\Delta^{\mu\nu}_{\alpha\beta}\nabla_{\gamma}(\Delta^{\alpha\beta}_{\sigma\delta}\rho^{\gamma\sigma\delta})+\frac{1}{7\beta_{\pi}}\pi^{\gamma<\mu}\pi^{\nu>\beta}\omega_{\gamma\beta}.
\end{align}

\renewcommand{\theequation}{B.\arabic{equation}}
	\section{ Evolution equation of $\rho^{\mu\nu\lambda}$ }\label{calculation}
	In this appendix, we derive the evolution equation of third rank tensor $\rho^{\mu\nu\lambda}$. In the process of deriving it, firstly we took the comoving derivative of $\rho^{\mu\nu\lambda}$ and obtain the following equation 
 \begin{align}\label{b1}
	\dot{\rho}^{<\mu\nu\lambda>}+\frac{1}{\tau_R}\rho^{\mu\nu\lambda}=&\,\underbrace{\Delta^{\mu\nu\lambda}_{\gamma\delta\sigma}\left[\dot{\Delta}_{\alpha\beta\rho}^{\gamma\delta\sigma}\int dP\frac{p^{\alpha}p^{\beta}p^{\rho}}{(u\cdot p)}\delta f\right]}_{(i)}-	\Delta_{\alpha\beta\rho}^{\mu\nu\lambda}
	\int dP~p^{\alpha}p^{\beta}p^{\rho}\Bigg[\underbrace{\frac{\dot{f_{\rm eq}}}{(u\cdot p)}}_{(ii)} \nonumber\\
	&+\underbrace{\frac{p^{\gamma}\nabla_{\gamma}f}{(u\cdot p)^2}}_{(iii)}+\underbrace{\frac{\delta f}{(u\cdot p)^2}D(u\cdot p)}_{(iv)}\Bigg].
\end{align}

\begin{center}
	\textbf{Term (i)}
\end{center}
The first term of Eq. \eqref{b1} is
\begin{equation}
(i) = \Delta^{\mu\nu\lambda}_{\gamma\delta\sigma}\left[\dot{\Delta}_{\alpha\beta\rho}^{\gamma\delta\sigma}\int dP\frac{p^{\alpha}p^{\beta}p^{\rho}}{(u\cdot p)}\delta f\right].
\end{equation}
The six rank tensor, $\Delta^{\mu\nu\lambda}_{\gamma\delta\sigma}$ is orthogonal to 4-velocity and second rank tensor, i.e. $\Delta^{\mu\nu\lambda}_{\gamma\delta\sigma}u_{\mu}=0$, $\Delta^{\mu\nu\lambda}_{\gamma\delta\sigma}\Delta_{\mu\nu}=0$. 

Using these properties and the form of second-order viscous correction, the first term contribution in terms of thermodynamic integral is obtained as
\begin{align}
	(i) = \Delta^{\mu\nu\lambda}_{\gamma\delta\sigma}\left[\dot{\Delta}_{\alpha\beta\rho}^{\gamma\delta\sigma}\int dP\frac{p^{\alpha}p^{\beta}p^{\rho}}{(u\cdot p)}\delta f\right]&=-\frac{3\beta}{\beta_{\pi}}I_{52}^{(2)}\left(\dot{u}^{<\mu}~\pi^{\nu\lambda>}\right)-\frac{6\tau_{\pi}\beta}{\beta_{\pi}}I_{52}^{(2)}~\left(\omega^{\gamma<\mu}\pi^{\nu}_{\gamma}\dot{u}^{\lambda>}\right) \nonumber\\
	&-\frac{15\beta}{7\beta_{\pi}^2}I_{52}^{(2)}~\left(\pi^{\gamma<\mu}\pi^{\nu}_{\gamma}\dot{u}^{\lambda>}\right)+\frac{2\beta\tau_{\pi}}{\beta_{\pi}}I_{52}^{(2)}~\theta~\left(\pi^{<\mu\nu}\dot{u}^{\lambda>}\right).
\end{align}	
The values of different thermodynamic integral can be found in Appendix \eqref{thermo_inte}.

\begin{center}
	\textbf{Term (ii)}	
\end{center}
The second term of Eq. \eqref{b1} is
	\begin{equation}
		(ii) = -\Delta_{\alpha\beta\rho}^{\mu\nu\lambda}\int dP \frac{p^{\alpha}p^{\beta}p^{\rho}}{(u\cdot p)}\dot{f_{\rm eq}}.
	\end{equation}
The comoving derivative of $\dot{f_{\rm eq}}$ is
\begin{equation}
	\dot{f_{\rm eq}}=-\left[\left(u\cdot p\right)\left(\frac{\beta\theta}{
	3}-\frac{\beta}{12P}\pi^{\mu\nu}\sigma_{\mu\nu}\right)+\beta p^{\mu}\dot{u}_{\mu}\right]f_{\rm eq}.
\end{equation}	
Since, $u^{\alpha}\Delta_{\alpha\beta\rho}^{\mu\nu\lambda}=u^{\beta}\Delta_{\alpha\beta\rho}^{\mu\nu\lambda}=u^{\rho}\Delta_{\alpha\beta\rho}^{\mu\nu\lambda}=\Delta^{\alpha\beta}\Delta_{\alpha\beta\rho}^{\mu\nu\lambda}=\Delta^{\alpha\rho}\Delta_{\alpha\beta\rho}^{\mu\nu\lambda}=\Delta^{\rho\beta}\Delta_{\alpha\beta\rho}^{\mu\nu\lambda}=0$. Hence, the second term has no contribution.
\begin{equation}
	(ii) = -\Delta_{\alpha\beta\rho}^{\mu\nu\lambda}\int dP \frac{p^{\alpha}p^{\beta}p^{\rho}}{(u\cdot p)}\dot{f_{\rm eq}}=0.
\end{equation}	
\begin{center}
	\textbf{Term (iii)}
\end{center}	

The third term of Eq. \eqref{b1} is	
\begin{align}
	(iii) = -\Delta_{\alpha\beta\rho}^{\mu\nu\lambda}\int dP~\frac{p^{\alpha}p^{\beta}p^{\rho}p^{\gamma}\nabla_{\gamma}(f_{\rm eq}+\delta f)}{(u\cdot p)^2},
\end{align}
which can be further simplified as
\begin{align}
	(iii) =-\Delta_{\alpha\beta\rho}^{\mu\nu\lambda}\int dP~p^{\alpha}p^{\beta}p^{\rho}p^{\gamma}&\Bigg[\nabla_{\gamma}\left(\frac{f_{\rm eq}}{(u\cdot p)^2}\right)-f_{\rm eq}\nabla_{\gamma}\left(\frac{1}{(u\cdot p)^2}\right)+\nabla_{\gamma}\left(\frac{\delta f}{(u\cdot p)^2}\right) \nonumber\\
	&-\delta f~\nabla_{\gamma}\left(\frac{1}{(u\cdot p)^2}\right)\Bigg],\\
	=-\Delta_{\alpha\beta\rho}^{\mu\nu\lambda}\int dP~p^{\alpha}p^{\beta}p^{\rho}p^{\gamma}&\Bigg[\underbrace{\nabla_{\gamma}\left(\frac{f_{\rm eq}}{(u\cdot p)^2}\right)}_{(a)}+\underbrace{\frac{2p^{\sigma}\left(\nabla_{\gamma}u_{\sigma}\right)}{(u\cdot p)^3}f_{\rm eq}}_{(b)}+\underbrace{\nabla_{\gamma}\left(\frac{\delta f}{(u\cdot p)^2}\right)}_{(c)} \nonumber\\
	&+\underbrace{\frac{2p^{\sigma}\left(\nabla_{\gamma}u_{\sigma}\right)}{(u\cdot p)^3}\delta f}_{(d)}\Bigg].
\end{align}

The term (a) is
\begin{equation}
	(a) = -\Delta_{\alpha\beta\rho}^{\mu\nu\lambda}~\nabla_{\gamma}\int dp~\frac{p^{\alpha}p^{\beta}p^{\rho}p^{\gamma}}{(u\cdot p)^2}f_{\rm eq}.
\end{equation}
Using the definition of thermodynamic integral, the above equation is obtained as
\begin{align}
		(a)=&-\Delta_{\alpha\beta\rho}^{\mu\nu\lambda}\Bigg[~I_{42}^{(2)}\bigg\{\nabla_{\gamma}\left(\Delta^{\alpha\beta}\right)\Delta^{\rho\gamma}+\nabla_{\gamma}\left(\Delta^{\rho\gamma}\right)\Delta^{\alpha\beta}+\nabla_{\gamma}\left(\Delta^{\alpha\rho}\right)\Delta^{\beta\gamma}+\nabla_{\gamma}\left(\Delta^{\beta\gamma}\right)\Delta^{\alpha\rho} \nonumber\\
	&+\nabla_{\gamma}\left(\Delta^{\rho\beta}\right)\Delta^{\alpha\gamma}+\nabla_{\gamma}\left(\Delta^{\alpha\gamma}\right)\Delta^{\rho\beta}\bigg\} + \left(\nabla_{\gamma}I_{42}^{(2)} \right) \bigg\{\Delta^{\alpha\beta}\Delta^{\rho\gamma} + \Delta^{\alpha\rho}\Delta^{\beta\gamma} + \Delta^{\alpha\gamma}\Delta^{\beta\rho} \bigg\} \Bigg],\\\nonumber
	=&-\Delta_{\alpha\beta\rho}^{\mu\nu\lambda}~I_{42}^{(2)}\left[\nabla^{\rho}\left(\Delta^{\alpha\beta}\right)+\nabla^{\beta}\left(\Delta^{\alpha\rho}\right)+\nabla^{\alpha}\left(\Delta^{\rho\beta}\right)\right],
\end{align}
where, 
\begin{equation}
	\nabla^{\rho}\left(\Delta^{\alpha\beta}\right) = \nabla^{\rho}\left(g^{\alpha\beta}-u^{\alpha}u^{\beta}\right)
	= \underbrace{-\left(\nabla^{\rho}u^{\alpha}\right)u^{\beta}-u^{\alpha}\left(\nabla^{\rho}u^{\beta}\right).}_{\text{vanishes on contraction with six rank tensor}}
\end{equation}
Hence, the contribution from (a) is zero.

The term (b) is
\begin{align}
	(b)=&-2\left(\nabla_{\gamma}u_{\sigma}\right)\Delta_{\alpha\beta\rho}^{\mu\nu\lambda}\int dP~\frac{p^{\alpha}p^{\beta}p^{\rho}p^{\gamma}p^{\sigma}}{(u\cdot p)^3}f_{\rm eq},\\\nonumber
	=&-2\left(\nabla_{\gamma}u_{\sigma}\right)\Delta_{\alpha\beta\rho}^{\mu\nu\lambda}\Bigg[I_{50}^{(3)}\left(u^{\alpha}u^{\beta}u^{\rho}u^{\gamma}u^{\sigma}\right)+I_{51}^{(3)}\Big(\Delta^{\alpha\beta}u^{\rho}u^{\gamma}u^{\sigma}+\Delta^{\alpha\rho}u^{\beta}u^{\gamma}u^{\sigma}+
\Delta^{\alpha\gamma}u^{\beta}u^{\rho}u^{\sigma}
\\\nonumber
&+\Delta^{\alpha\sigma}u^{\beta}u^{\rho}u^{\gamma}+\Delta^{\beta\rho}u^{\alpha}u^{\gamma}u^{\sigma}+\Delta^{\beta\mu}u^{\alpha}u^{\rho}u^{\sigma}+\Delta^{\beta\sigma}u^{\alpha}u^{\rho}u^{\gamma}+	\Delta^{\rho\gamma}u^{\alpha}u^{\beta}u^{\sigma}+\Delta^{\rho\sigma}u^{\alpha}u^{\beta}u^{\gamma}\\\nonumber
&+\Delta^{\gamma\sigma}u^{\alpha}u^{\beta}u^{\rho}\Big)+I_{52}^{(3)}\Big(\Delta^{\alpha\beta}\Delta^{\rho\gamma}u^{\sigma}+\Delta^{\alpha\rho}\Delta^{\beta\gamma}u^{\sigma}+
	\Delta^{\alpha\gamma}\Delta^{\beta\rho}u^{\sigma}+
	\Delta^{\alpha\beta}\Delta^{\rho\sigma}u^{\gamma}
	+\Delta^{\alpha\rho}\Delta^{\beta\sigma}u^{\gamma}\\\nonumber
&+\Delta^{\alpha\sigma}\Delta^{\rho\beta}u^{\gamma}+\Delta^{\alpha\beta}\Delta^{\gamma\sigma}u^{\rho}+\Delta^{\alpha\gamma}\Delta^{\beta\sigma}u^{\rho}+
\Delta^{\alpha\beta}\Delta^{\gamma\sigma}u^{\rho}+\Delta^{\alpha\gamma}\Delta^{\beta\sigma}u^{\rho}+
\Delta^{\alpha\beta}\Delta^{\gamma\sigma}u^{\rho}\\\nonumber
&+\Delta^{\alpha\sigma}\Delta^{\beta\gamma}u^{\rho}+\Delta^{\alpha\rho}\Delta^{\gamma\sigma}u^{\beta}+\Delta^{\alpha\gamma}\Delta^{\rho\sigma}u^{\beta}+
\Delta^{\alpha\sigma}\Delta^{\rho\gamma}u^{\beta}+\Delta^{\beta\rho}\Delta^{\gamma\sigma}u^{\alpha}+
\Delta^{\beta\gamma}\Delta^{\rho\sigma}u^{\alpha}\\
&+\Delta^{\beta\sigma}\Delta^{\rho\gamma}u^{\alpha}\Big)\Bigg].
\end{align}
On contraction with six rank tensor, this term also vanishes.\\

The term (c) is
\begin{align}
	(c)&=-\Delta_{\alpha \beta \rho}^{\mu\nu\lambda}\int dP~p^{\alpha}p^{\beta}p^{\rho}p^{\gamma}\nabla_{\gamma}\left(\frac{\delta f}{(u\cdot p)^2}\right),\\
	&=-\Delta_{\alpha \beta \rho}^{\mu\nu\lambda}\nabla_{\gamma}\int dP~p^{\alpha}p^{\beta}p^{\rho}p^{\gamma}\left(\frac{\delta f}{(u\cdot p)^2}\right).\label{c}
\end{align}

Using the following identity,
	\begin{equation}\label{identity}
	\Delta^{\alpha\rho}~p_{\rho}=p^{<\alpha>}=p^{\alpha}-(u\cdot p)u^{\alpha},
\end{equation}
 Eq. \eqref{c} is obtained as

\begin{align}
  (c)  =&-\Delta^{\mu\nu\lambda}_{\alpha\beta\rho}(\nabla_{\mu}u^{\alpha})\!\int\! dP ~p^{\beta}p^{\rho}p^{\mu}~\frac{\delta f}{u\cdot p}-\Delta^{\mu\nu\lambda}_{\alpha\beta\rho}\!\int\! dP u^{\alpha}p^{\beta}p^{\rho}p^{\mu}\,\nabla_{\mu}\!\left(\frac{\delta f}{u\cdot p}\right)-\Delta^{\mu\nu\lambda}_{\alpha\beta\rho}\,\nabla_{\mu}\!\int\! dP p^{<\alpha>}p^{\beta}p^{\rho}p^{\mu}\frac{\delta f}{(u\cdot p)^2},\nonumber\\
  =&-\Delta^{\mu\nu\lambda}_{\alpha\beta\rho}(\nabla_{\mu}u^{\alpha})\int dP ~p^{\beta}p^{\rho}p^{\mu}~\frac{\delta f}{u\cdot p}-\Delta^{\mu\nu\lambda}_{\alpha\beta\rho}(\nabla_{\mu}u^{\beta})\int dP ~p^{\alpha}p^{\rho}p^{\mu}~\frac{\delta f}{u\cdot p} -\Delta^{\mu\nu\lambda}_{\alpha\beta\rho}(\nabla_{\mu}u^{\rho})\int dP ~p^{\alpha}p^{\beta}p^{\mu}~\frac{\delta f}{u\cdot p}\nonumber\\
  &-\Delta^{\mu\nu\lambda}_{\alpha\beta\rho}(\nabla_{\mu}u^{\mu})\int dP ~p^{\alpha}p^{\beta}p^{\rho}~\frac{\delta f}{u\cdot p}-\Delta^{\mu\nu\lambda}_{\alpha\beta\rho}\nabla_{\mu}\int dP p^{<\alpha>}p^{<\beta>}p^{<\rho>}p^{<\mu>}\left(\frac{\delta f}{(u\cdot p)^2}\right).
\end{align}
After putting the form of second order viscous correction to the distribution function, the above equation comes out to be
\begin{align}
	(c) =&-\frac{18\beta\tau_{\pi}}{5\beta_{\pi}}\dot{u}^{\gamma}\Delta^{\mu\nu\lambda}_{\alpha\beta\rho}I_{42}^{(1)}\left[(\nabla^{\beta}u^{\alpha})\pi^{\rho}_{\gamma}+(\nabla^{\rho}u^{\alpha})\pi^{\beta}_{\gamma}\right]+\frac{3\beta\tau_{\pi}}{5\beta_{\pi}}I_{42}^{(1)}\Delta^{\mu\nu\lambda}_{\alpha\beta\rho}\left[(\nabla^{\beta}u^{\alpha})(\nabla^{\gamma}\pi^{\rho}_{\gamma})+(\nabla^{\rho}u^{\alpha})(\nabla^{\gamma}\pi^{\beta}_{\gamma})\right]\nonumber\\
 &-\frac{9\beta\tau_{\pi}}{\beta_{\pi}}\Delta^{\mu\nu\lambda}_{\alpha\beta\rho}I_{63}^{(3)}\left[4(\nabla^{\beta}u^{\alpha})\pi^{\rho}_{\gamma}\dot{u}^{\gamma}+4(\nabla^{\gamma}u^{\alpha})\pi^{\beta}_{\gamma}\dot{u}^{\rho}+2(\nabla^{\gamma}u^{\alpha})\pi^{\rho\beta}\dot{u}_{\gamma}\right]+\frac{3\beta\tau_{\pi}}{\beta_{\pi}}I_{63}^{(3)}\Delta^{\mu\nu\lambda}_{\alpha\beta\rho}\Big[4(\nabla^{\beta}u^{\alpha})(\nabla^{\gamma}\pi^{\rho}_{\gamma})\nonumber\\
 &+4(\nabla^{\gamma}u^{\alpha})(\nabla^{\rho}\pi^{\beta}_{\gamma})+2(\nabla^{\gamma}u^{\alpha})(\nabla_{\gamma}\pi^{\rho\beta})
\Big]-\rho^{\mu\nu\lambda}\theta+\frac{3}{7}\nabla^{<\mu}\pi^{\nu\lambda>}+6\Delta^{\mu\nu\lambda}_{\alpha\beta\rho}~\pi_{\gamma}^{\beta}\omega^{\rho\gamma}\nabla^{\alpha}\left(\frac{\beta\tau_{\pi}}{\beta_{\pi}}I_{63}^{(3)}\right)\nonumber\\
&+\frac{\beta\tau_{\pi}}{\beta_{\pi}}I_{63}^{(3)}\Delta_{\alpha\beta\rho}^{\mu\nu\lambda}\left[6~\omega^{\rho\gamma}\nabla^{\alpha}\pi_{\gamma}^{\beta}+6\pi_{\gamma}^{\beta}\nabla^{\alpha}\omega^{\rho\gamma}\right]-\frac{30}{14}\Delta^{\mu\nu\lambda}_{\alpha\beta\rho}\pi^{\gamma\beta}\pi^{\rho}_{\gamma}~\nabla^{\alpha}\left(\frac{\beta}{\beta_{\pi}^2}I_{63}^{(3)}\right)-\frac{30\beta}{7\beta_{\pi}^2}I_{63}^{(3)}\nabla^{\mu\nu\lambda}_{\alpha\beta\rho}\pi^{\rho}_{\gamma}\nabla^{\alpha}\pi^{\gamma\beta}\nonumber\\
&+2\pi^{\beta\rho}\theta\Delta^{\mu\nu\lambda}_{\alpha\beta\rho}\nabla^{\alpha}\left(\frac{\tau_{\pi}\beta I_{63}^{(3)}}{\beta_{\pi}}\right)+\frac{2\beta\tau_{\pi}}{\beta_{\pi}}\theta I_{63}^{(3)}\Delta^{\mu\nu\lambda}_{\alpha\beta\rho}\nabla^{\alpha}\pi^{\beta\rho}-\Delta_{\alpha\beta\rho}^{\mu\nu\lambda}\left(10\Delta^{\rho\gamma}\pi^{\alpha\sigma}\pi^{\beta}_{\sigma}+8\pi^{\rho\alpha}\pi^{\beta\gamma}\right)\nabla_{\mu}\left(\frac{\beta^2 I_{84}^{(4)}}{4\beta_{\pi}^2}\right)\nonumber\\
&-\Delta^{\mu\nu\lambda}_{\alpha\beta\rho}\left(\frac{\beta^2 I_{84}^{(4)}}{4\beta_{\pi}^2}\right)\nabla_{\gamma}\left(10\Delta^{\rho\gamma}\pi^{\alpha\sigma}\pi^{\beta}_{\sigma}+8\pi^{\rho\alpha}\pi^{\beta\gamma}\right)-\Delta_{\alpha\beta\rho}^{\mu\nu\lambda}\left(10\Delta^{\rho\gamma}\pi^{\alpha\sigma}\pi^{\beta}_{\sigma}+8\pi^{\rho\alpha}\pi^{\beta\gamma}\right)\nabla_{\mu}\left(\frac{\beta I_{84}^{(5)}}{4\beta_{\pi}^2}\right)\nonumber\\
&-\Delta^{\mu\nu\lambda}_{\alpha\beta\rho}\left(\frac{\beta I_{84}^{(5)}}{4\beta_{\pi}^2}\right)\nabla_{\gamma}\left(10\Delta^{\rho\gamma}\pi^{\alpha\sigma}\pi^{\beta}_{\sigma}+8\pi^{\rho\alpha}\pi^{\beta\gamma}\right).
\end{align}

The term (d) is 

\begin{equation}
	(d) = -\Delta_{\alpha \beta \rho}^{\mu\nu\lambda}\int dP~p^{\alpha}p^{\beta}p^{\rho}p^{\gamma}\left\{\frac{2p^{\sigma}\left(\nabla_{\gamma}u_{\sigma}\right)}{(u\cdot p)^3}\right\}\delta f.
 \end{equation}
 Putting the form of $\delta f$ and solving for its contribution, we have
 \begin{align}
	(d)=&-\frac{\beta}{\beta_{\pi}}\left(\frac{24\tau_{\pi}}{5}\right)I_{63}^{(3)}~\dot{u}^{\rho}\left[3\sigma^{<\mu\nu}\pi^{\lambda>}_{\rho}\right]+\frac{12\beta\tau_\pi}{5\beta_\pi}I_{63}^{(3)}\Delta_{\alpha\beta\rho}^{\mu\nu\lambda}(\sigma^{\alpha\beta}\nabla^{\gamma}\pi^{\rho}_{\gamma})-6\tau_{\pi}\left(\frac{\beta}{\beta_{\pi}}\right)I_{84}^{(5)}\Big[2\sigma^{<\lambda\mu}\pi^{\nu>}_{\rho}\dot{u}^{\rho}
	+2\sigma^{<\lambda\nu}\pi^{\mu>}_{\rho}\dot{u}^{\rho}\nonumber\\
 &+2\omega^{<\lambda\mu}\pi^{\nu>}_{\rho}\dot{u}^{\rho}+2\omega^{<\lambda\nu}\pi^{\mu>}_{\rho}\dot{u}^{\rho}+2\sigma^{<\lambda}_{~\gamma}\pi^{\mu\nu>}\dot{u}^{\gamma}+2\sigma_{\gamma}^{~<\lambda}\pi^{\mu\nu>}\dot{u}^{\gamma}
	+\frac{10}{3}~\theta~\pi^{<\mu\nu}\dot{u}^{\lambda>} 
 +2~\theta~\pi^{<\mu\lambda}\dot{u}^{\nu>}+2~\theta~\pi^{<\nu\lambda}\dot{u}^{\mu>}\Big]\nonumber\\
 &+\tau_{\pi}\left(\frac{\beta}{\beta_{\pi}}\right)I_{84}^{(5)}\Delta^{\mu\nu\lambda}_{\alpha\beta\rho}\Big[2(\nabla^{\gamma}\pi^{\beta\alpha})(\nabla^{\rho}u_{\gamma})+2(\nabla^{\gamma}\pi^{\alpha\beta})(\nabla_{\gamma}u^{\rho})+6\theta(\nabla^{\rho}\pi^{\alpha\beta})\Big].
\end{align}		

\begin{center}
	\textbf{Term (iv)}
\end{center}
The term (iv) is
\begin{equation}
	(iv) = -\Delta_{\alpha \beta \rho}^{\mu\nu\lambda}\int dP~p^{\alpha}p^{\beta}p^{\rho}\left[\frac{\delta f}{(u\cdot p)^2}D(u\cdot p)\right].
\end{equation}	
Simplifying the above term and using the form of $\delta f$, the term (iv) is obtained as
\begin{align}
	(iv) =&\,-\Delta_{\alpha\beta\rho}^{\mu\nu\lambda}~\int dP~ \frac{p^{\alpha}p^{\beta}p^{\rho}p^{\gamma}\dot{u}_{\gamma}}{(u\cdot p)^2}\delta f,\\
	=&-3\frac{\beta}{\beta_{\pi}}I_{63}^{(3)}~\pi^{<\mu\nu}\dot{u}^{\lambda>}+6\beta\frac{\tau_{\pi}}{\beta_{\pi}}I_{63}^{(3)}~\dot{u}^{<\mu}\pi^{\nu}_{\gamma}\omega^{\lambda>\gamma}-\frac{30\beta}{14\beta_{\pi}^2}I_{63}^{(3)}~\dot{u}^{<\mu}\pi_{\gamma}^{~\nu}\pi^{\lambda>\gamma}+
	\frac{2\beta\tau_{\pi}}{\beta_{\pi}}I_{63}^{(3)}\theta\left(\dot{u}^{<\mu}\pi^{\nu\lambda>}\right) \nonumber\\
	&-\frac{\beta^2}{4\beta_{\pi}^2}I_{84}^{(4)}\left[10\left(\pi^{\gamma<\mu}\pi^{\nu}_{\gamma}\dot{u}^{\lambda>}+8\dot{u}_{\gamma}\pi^{<\mu\nu}\pi^{\lambda>\gamma}\right)\right]-\frac{\beta}{4\beta_{\pi}^2}I_{84}^{(5)}\left(10~\pi^{\gamma<\mu}\pi^{\nu}_{~\gamma}\dot{u}^{\lambda>}+8\dot{u}_{\gamma}\pi^{<\mu\nu}\pi^{\lambda>\gamma}\right).
\end{align}	
Finally, adding the terms (i), (ii), (iii) and (iv) and using the values of different integrals given in Appendix \eqref{thermo_inte}, the evolution equation for $\rho^{\mu\nu\lambda}$ is obtained as,
\begin{align}
	\dot{\rho}^{<\mu\nu\lambda>}+\frac{1}{\tau_\rho}\rho^{\mu\nu\lambda}=&~
	\frac{3}{7}\nabla^{<\mu}\pi^{\nu\lambda>}-~\frac{18}{7}\dot{u}^{<\mu}\pi^{\nu\lambda>}-\frac{187}{81}\rho^{\mu\nu\lambda}\theta-\frac{10}{7}\tau_{\pi}\dot{u}^{<\mu}\pi^{\nu\lambda>}\theta-\frac{36}{7}\tau_{\pi}\omega^{\gamma<\mu}\pi^{\nu}_{\gamma}\dot{u}^{\lambda>}\nonumber\\
 &-\frac{389}{441\beta_{\pi}}\pi^{\gamma<\mu}\pi^{\nu}_{\gamma}\dot{u}^{\lambda>}-\frac{343}{105\beta_{\pi}}\dot{u}_{\gamma}\pi^{\gamma<\mu}\pi^{\nu\lambda>}+\frac{18}{7}\tau_{\pi}\dot{u}_{\gamma}\omega^{\gamma<\mu}\pi^{\nu\lambda>}-\frac{6}{7}\tau_{\pi}\omega^{\gamma<\mu}\nabla_{\gamma}\pi^{\nu\lambda>} \nonumber\\
& -\frac{6}{7}\tau_{\pi}\omega^{\gamma<\mu}\nabla^{\nu}\pi^{\lambda>}_{\gamma}-\frac{6}{7}\tau_{\pi}\pi_{\gamma}^{<\mu}\nabla^{\nu}\omega^{\lambda>\gamma}-\frac{47}{63\beta_{\pi}}\pi^{<\mu\nu}\nabla_{\gamma}\pi^{\lambda>\gamma}-\frac{11}{21\beta_{\pi}}\pi^{\gamma<\mu}\nabla_{\gamma}\pi^{\nu\lambda>} \nonumber\\
&-\frac{665}{441\beta_{\pi}}\pi^{\gamma<\mu}\nabla^{\nu}\pi^{\lambda>}_{\gamma}.
\end{align}

\section{Thermodynamic integrals}\label{thermo_inte}

We define the thermodynamic integrals
\begin{equation}
I_{n q}^{(m)}=\frac{1}{(2 q+1) ! !} \int d p(u \cdot p)^{n-2 q-m}\left(\Delta_{\alpha \beta} p^\alpha p^\beta\right)^q f_{\rm eq},
\end{equation}
and state the following relation
\begin{equation}
I_{n q}^{(0)}=\frac{1}{\beta}\left[-I_{n-1, q-1}^{(0)}+(n-2 q) I_{n-1, q}^{(0)}\right] \text {. }
\end{equation}
With the above definition,  we identify $I_{20}^{(0)}=\epsilon$ and $I_{21}^{(0)}=-P$. The values of different integrals are as follows\\
 \begin{center}
 $I_{52}^{(2)}=\frac{4T^5}{5\pi^2}=I_{42}^{(1)}$, $I_{84}^{(4)}=\frac{4T^6}{63\pi^2}$,  $I_{84}^{(5)}=\frac{4T^5}{63\pi^2}$, $I_{63}^{(3)}=-\frac{4T^5}{35\pi^2}$.
 \end{center}
\end{appendix}

\bibliography{references}

\end{document}